\documentstyle[fleqn,espcrc2,epsf]{article}

\newcommand{\lsim}{\mathrel{\raisebox{-.6ex}{$\stackrel{\textstyle<}{\sim}$}}}

\begin{document}
\pagestyle{plain}

\title{\vspace{-.7in}
\font\fortssbx=cmssbx10 scaled \magstep1
\hbox to \hsize{
\hbox{\fortssbx University of Wisconsin - Madison}
\hfill$\vtop{\hbox{\normalsize\bf MADPH-97-1005}
                \hbox{\normalsize July 1997}
                 \hbox{\hfil}}$ }
Higgs Searches at Future Colliders:\\
Potentials for Discovery and Diagnosis\thanks{Invited review talk at {\it The 5th International Conference on Supersymmetries in Physics (SUSY\,97)}, Philadelphia, Pennsylvania, May 27--31, 1997.}}

\author{V. Barger\address{Physics Department, University of Wisconsin, Madison, WI 53706}}

\begin{abstract}

The prospects for Higgs boson discovery and study at present and future colliders are reviewed, with emphasis on the Minimal Supersymmetric Standard Model (MSSM) states. The expected experimental coverages with different production and decay modes are outlined. Particular attention is given to the Higgs couplings predicted when the top quark Yukawa coupling has an infrared fixed point. Characteristic Higgs boson mass spectra from the minimal supergravity (mSUGRA) model are considered and the decoupling/non-decoupling regimes of the lightest Higgs boson are discussed. General tests are described to ascertain whether a discovered light Higgs boson is a SM or MSSM state and to diagnose its fundamental properties. The constraints on the Higgs mass spectra implied by an interesting cosmological relic density of the lightest supersymmetric particle are described.

\end{abstract}

\maketitle\thispagestyle{empty}

\section{Introduction}
\noindent
``What breaks the electroweak symmetry?" is the central physics issue of our time. Higgs bosons are the cornerstones of our theoretical models. The crucial tests of the spontaneous symmetry breaking (EWSB) mechanism are the Higgs boson interactions with the $W$ and $Z$ gauge fields. 

There are numerous options for Higgs multiplets. The Standard Model (SM) assumes one SU(2) doublet of Higgs fields and has one physical Higgs boson ($h_{\rm SM}^0$). The Minimal Supersymmetric Standard Model (MSSM) contains two doublets of Higgs fields and five physical Higgs bosons ($h^0$, $H^0$, $A^0$, $H^\pm$). Non-supersymmetric two-Higgs doublet models have a phase in the vacuum expectation values which is interesting for CP violation (e.g., a muon electric dipole moment). Higher dimensional Higgs multiplets such as $I=3/2$ contain doubly charged Higgs fields. If no neutral Higgs bosons are found to exist with mass $m_h \lsim 0.8$~TeV, then strong scattering must occur in the $WW$ sector. The thrust of this review will be the SM and MSSM Higgs scenarios. The reader is referred to other comprehensive recent reviews (\cite{haber,gunion,higgs-lep2,dawson} and references therein) for more thorough discussions.

\section{Indirect $m_{h_{\rm SM}^0}$ determinations}

Precision $M_W$ and $m_t$ measurements test radiative corrections through the relation
\begin{equation}
M_W = M_Z \left[ 1 - {\pi\alpha\over \sqrt 2 G_\mu M_W^2 (1-\delta r)} \right]^{1/2} \,.
\end{equation}
In the SM the loop  corrections $\delta r$ depend on $m_t^2$ and $\log m_h$ (see e.g.\ Ref.~\cite{langacker}). The optimal relative precision for tests of this relation are
\begin{equation}
\Delta M_W \approx {1\over140} \Delta m_t \,,
\end{equation}
and hence the $\Delta M_W$ accuracy is most critical. The estimated achievable accuracies $\Delta M_W$ at future machines with integrated luminosities $L$ are listed in Table~1.
\begin{table*}[t]
\centering
\caption[]{Estimated accuracy of $W$-mass measurements at LEP-2, Tevatron, Large Hadron Collider, Next Linear Collider ($\sqrt s = 0.5$~TeV), and First Muon Collider ($\sqrt s = 0.5$~TeV). See Ref.~\cite{haber}.}
\medskip
\begin{tabular}{|l|c|c|cc|c|c|c|}
\hline
& \multicolumn{2}{c|}{LEP-2}& \multicolumn{2}{c|}{Tevatron}& LHC& NLC ($e^+e^-$)& FMC ($\mu^+\mu^-$)\\
\hline
$L$ (fb$^{-1}$)& 0.1& 2& 2& 10& 10& 50& 50\\
$\Delta M_W$ (MeV)& 144& 34& 35& 20& 15& 15& 9\\
\hline
\end{tabular}
\end{table*}
Note that a luminosity at the LHC higher than 10~fb$^{-1}$ may not further improve $\Delta M_W$. As Fig.~1 shows, precise $\Delta M_W$ measurement will give pinpoint accuracy on $m_{h^0_{\rm SM}}$. A value $\Delta M_W = 6$~MeV (achievable at a muon collider with 100~fb$^{-1}$ luminosity\cite{mwmt}) would determine $m_{h^0_{\rm SM}}$ to
\begin{equation}
\Delta m_{h^0_{\rm SM}} = \pm 10 \left(m_h\over 100\rm\ GeV\right)\rm\ GeV\,.
\end{equation}
New physics contributions to $\delta r$ would likewise be stringently tested. An improved value of $\alpha(M_Z)$ from new measurements of $\sigma(e^+e^-\to\rm hadrons)$ at low energy is necessary to fully utilize a very accurate $M_W$ measurement. 

\begin{figure}[t]
\centering
\hspace{0in}\epsfxsize=7.5cm\epsffile{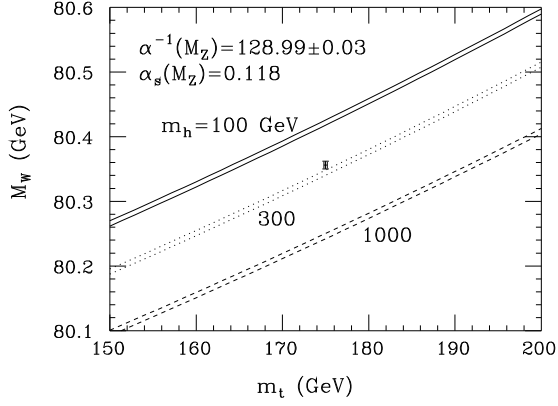}

\caption{Correlation between $\protect M_W$ and $\protect m_t$
in the SM with QCD and electroweak corrections
for $\protect m_h=100, 300$ and 1000~GeV. The data point
and error bars illustrate the possible accuracy for
the indirect $m_h$ determination assuming
$M_W=80.356\pm 0.006$ GeV and $m_t=175\pm 0.2$ GeV.
The widths of the bands indicate the uncertainty
in $\protect \alpha(M_Z)$. From Ref.~\protect\cite{mwmt}.}
\end{figure}

\section{SM Higgs -- the benchmark}

The Higgs mass $m_{h^0_{\rm SM}}$ is the only parameter of the SM Higgs sector. SM Higgs production occurs through the processes of Fig.~2. QCD radiative corrections have been evaluated for all hadronic production processes but $t\bar t h$ and $b\bar bh$. The SM backgrounds to the Higgs signals are well studied.

\begin{figure}[t]
\centering
\hspace{0in}\epsfxsize=7.5cm\epsffile{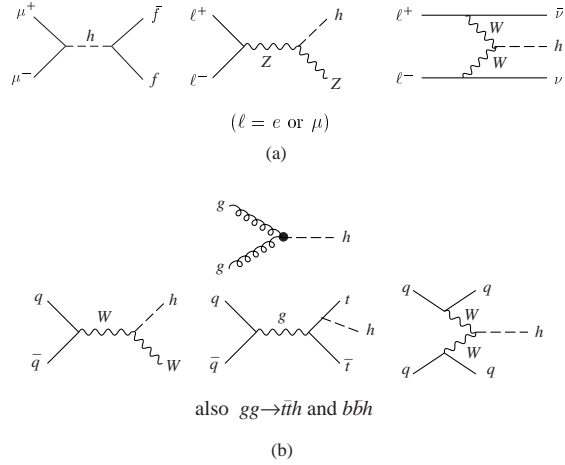}

\vspace{-.2in}  

\caption{Higgs boson production diagrams for (a)~$\ell^+\ell^-$ and (b)~$p\bar p, pp$ colliders.}
\end{figure}

The diagrams for the SM Higgs decays of primary interest are shown in Fig.~3.
The partial widths for Higgs boson decays to fermions are proportional to $m_f^2$, so a light SM Higgs decays dominantly to $b\bar b$, with $b\bar b$ branching fraction ${\sim}90\%$. The rare $\gamma\gamma$ mode with branching fraction ${\sim}10^{-3}$ for $m_{h^0_{\rm SM}} \sim 80$--140~GeV is important for LHC searches. Sophisticated detectors are required with $b$-flavor tagging capabilities for $b\bar b$ decays and high EM calorimeter resolution (${\sim}1\%$) for $\gamma\gamma$ detection. For heavy SM Higgs searches via the mode $h\to ZZ\to 4\ell$ (with $\ell = e$ or $\mu$), large angular coverage and muon detection are needed.

\begin{figure}[t]
\centering
\hspace{0in}\epsfxsize=7.5cm\epsffile{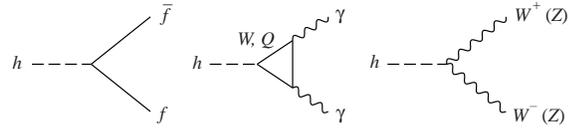}

\vspace{-.1in}

\caption{Higgs boson decay diagrams}
\end{figure}

In the ongoing search at LEP-2 for the SM Higgs, the collider energy is the major limiting factor. Figure~4 shows the potential $5\sigma$  discovery limits for $m_{h^0_{\rm SM}}$ at $\sqrt s = 175$, 192 and 205~GeV versus the minimum luminosity needed per experiment (with results from 4 experiments combined). The ALEPH collaboration has placed the bound\cite{aleph}
\begin{equation}
m_{h_{\rm SM}} > 70.7\rm\ GeV
\end{equation}
from 21.3~pb$^{-1}$ of luminosity in LEP-2 running up to the current energy $\sqrt s = 172$~GeV.

\begin{figure}[h]
\centering
\hspace{0in}\epsfxsize=7.5cm\epsffile{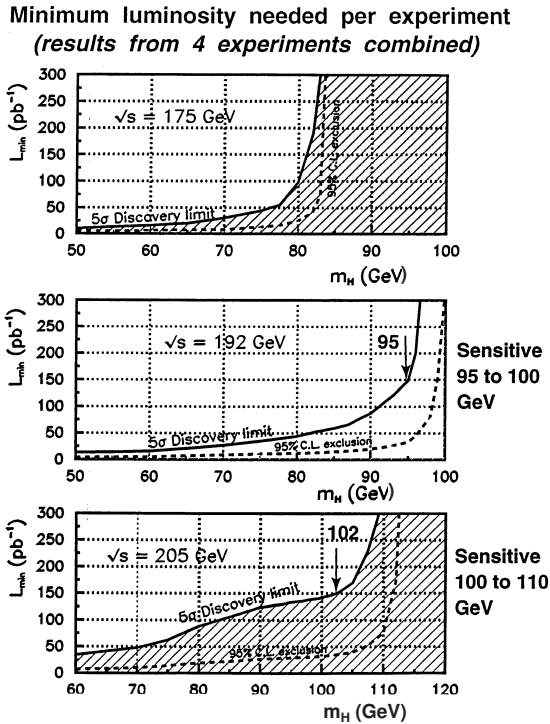}

\vspace{-.2in}

\caption{Minimum luminosity needed per experiment, in pb$^{-1}$, for a combined $5\sigma$ discovery (full line) or 95\% exclusion (dashed line) as a function of Higgs boson mass, at three center-of-mass energies. From Ref.~\protect\cite{higgs-lep2}.}
\end{figure}

Upgrades of the Tevatron (Main Injector, TeV-33) will allow for a SM Higgs search up to  $m_{h_{\rm SM}}\sim 120$~GeV\cite{yao} and possibly higher\cite{mrenna}, through the $Wh_{\rm SM}$ production mode with $W\to \ell\nu$ and $h_{\rm SM}\to b\bar b$ decays. The $5\sigma$ discovery requirements on luminosity for the detection are shown in Fig.~5. 

\begin{figure}[h]
\centering
\hspace{0in}\epsfxsize=7.5cm\epsffile{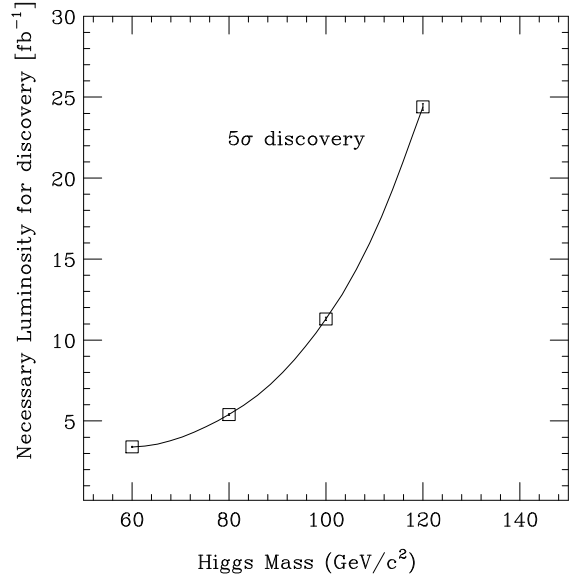}

\caption{Luminosity requirements for $5\sigma$ discovery of a SM Higgs boson at the Tevatron in $Wh_{\rm SM}$ production. From Ref.~\protect\cite{yao}. }
\end{figure}

Table~2 summarizes the discovery reach for SM Higgs searches at present and future colliders. The coverage by various modes at the different colliders is illustrated graphically in Fig.~6. Note that the combined LEP-2 + Tevatron Main Injector + LHC searches will cover the entire interesting $m_{h_{\rm SM}}$ range up to ${\cal O}$(TeV). 

\begin{table}
\caption{Discovery reach for SM Higgs bosons at present and future colliders.}
\centering
\hspace{0in}\epsfxsize=7.5cm\epsffile{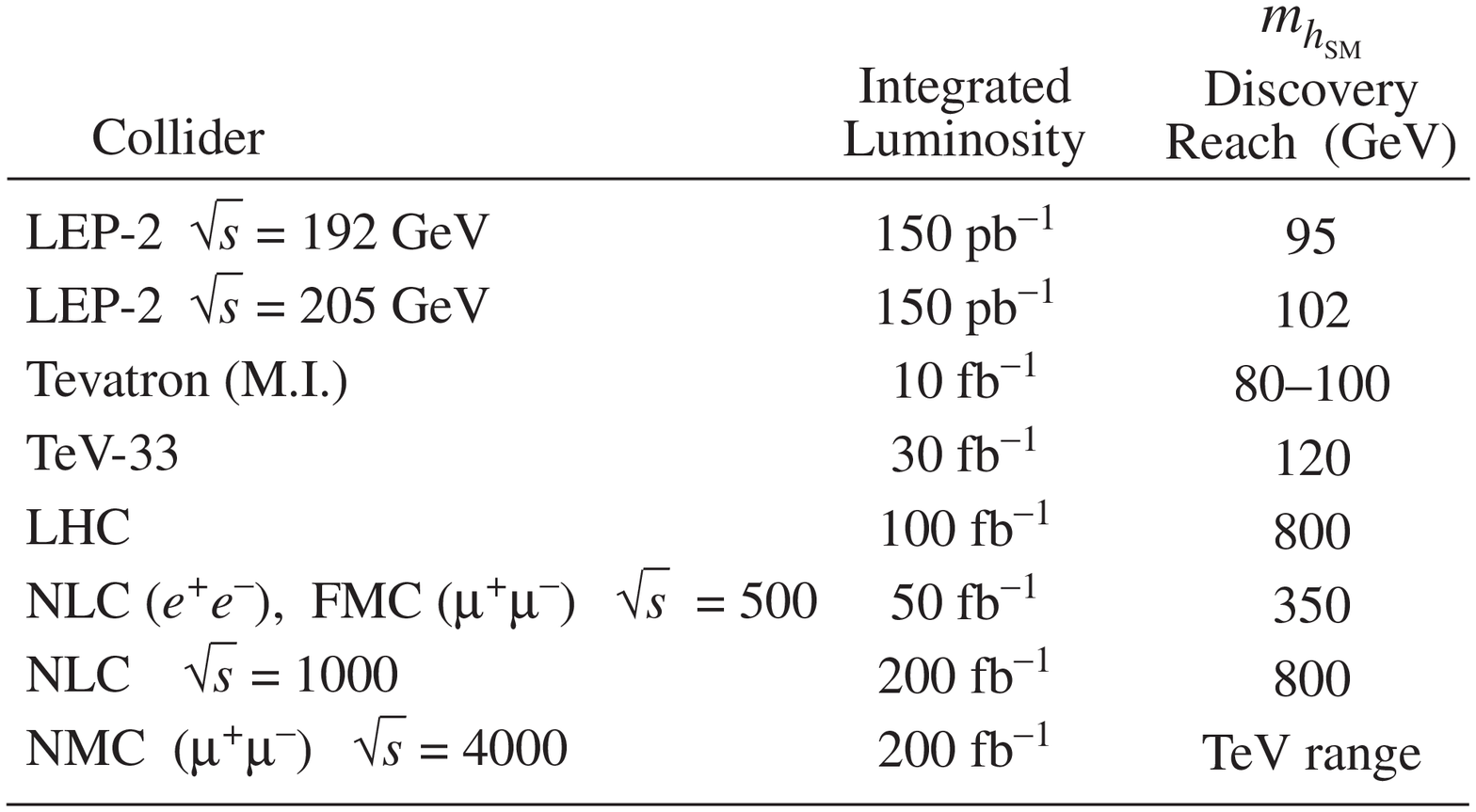}

\end{table}

\begin{figure}[t]
\centering
\hspace{0in}\epsfxsize=7.5cm\epsffile{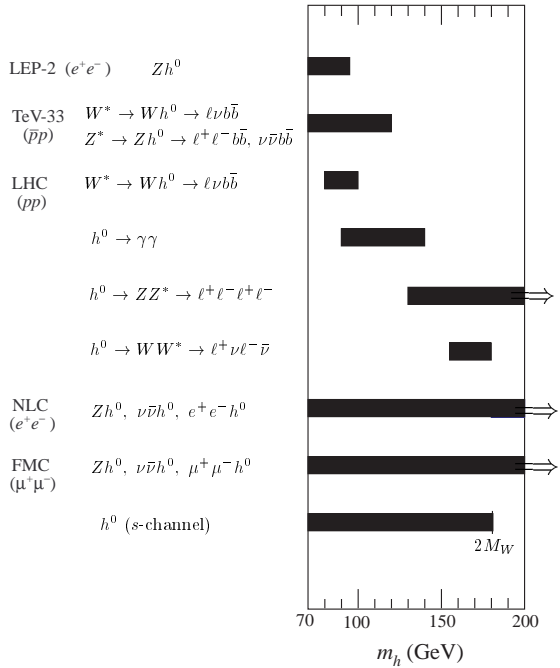}

\vspace{-.2in}

\caption[]{SM Higgs mass ranges over which statistically significant signals are expected in leading decay modes at future colliders (see Ref.~\cite{haber}).}
\end{figure}

\section{MSSM Higgs $(h^0, H^0, A^0, H^\pm)$ -- the target}

In the supersymmetric scenario there is an upper bound on the mass of the lightest Higgs boson. At tree level this bound is (see e.g.\ Ref.~\cite{hhg})
\begin{equation}
m_{h^0} \leq M_Z |\cos2\beta| \,,
\end{equation}
where $\tan\beta = v_u/v_d$ is the ratio of vacuum expectation values.
Radiative corrections can substantially raise the tree level bound to\cite{haber,gunion,higgs-lep2} 
\begin{equation}
m_{h^0} \lsim 130\rm\ GeV\,.
\end{equation}
The value of $m_{h^0}$ depends on $\tan\beta$, the stop masses and the stop mixing. In the large $m_A$ limit the radiatively corrected $m_h$ is given by\cite{carena}
\begin{eqnarray}
&&\hspace{-.25in} m_h^2 = M_Z^2 \cos^22\beta \left( 1 - {3m_t^2 t\over 8\pi^2 v^2}\right) \nonumber\\
&&\hspace{-.25in}{}+ {3m_t^4\over 4\pi^2v^2} \left[ t + {\tilde\kappa\over 2} + \left( {3\over 32\pi^2} {m_t^2\over v^2} - {2\alpha_3\over \pi}\right) \left(\tilde\kappa t + t^2\right) \right] \nonumber\\
\end{eqnarray}
where
\begin{eqnarray}
t &=& \ln \left( M_s^2\over m_t^2 \right) \,,\nonumber \\
\kappa &=& {2\tilde A_t^2\over M_S^2} \left( 1 - {\tilde A_t^2\over 12M_S^2}\right) \,,\nonumber \\
\tilde A_t &=& A_t + \mu/\tan\beta \,, \nonumber \\
M_S^2 &\simeq& m_{\tilde t_L}^2 +  m_{\tilde t_R}^2 \,.\nonumber
\end{eqnarray}
In a non-minimal supersymmetric model with extra Higgs doublets and singlets, the mass of at least one Higgs boson satisfies\cite{kolda}
\begin{equation}
m_{h^0} \lsim 150\rm\ GeV
\end{equation}
so long as the couplings remain perturbative to the Planck scale.

The current LEP-2 lower bounds on the Higgs masses in the MSSM are\cite{aleph}
\begin{eqnarray}
m_{h^0} &>& 62.5\rm\ GeV\,,\\
m_{A^0} &>& 62.5\rm\ GeV\,,\\
m_{H^\pm} &>& 44\rm\ GeV \,.
\end{eqnarray}
These limits are based on searches for the final states
\begin{eqnarray}
H^0 Z(hZ) &\to& b\bar b q\bar q, b\bar b\ell\bar\ell, b\bar b \nu\bar\nu, \tau\bar \tau q\bar q\,,\\
hA &\to& b\bar bb\bar b, \tau\bar\tau b\bar b\,.
\end{eqnarray}
The production cross sections have the $\beta$-dependences
\begin{eqnarray}
\sigma_{hZ}, \sigma_{HA} &\propto& \sin^2(\beta-\alpha)\,,\\
\sigma_{HZ}, \sigma_{hA} &\propto& \cos^2(\beta-\alpha)\,,
\end{eqnarray}
where $\alpha$ is the Higgs mixing angle. Figure~7 shows the regions of the $(\tan\beta, m_h)$ space presently excluded by ALEPH searches and theoretical constraints, with $M_S = 1$~TeV assumed.

\begin{figure}[t]
\centering
\hspace{0in}\epsfxsize=7.5cm\epsffile{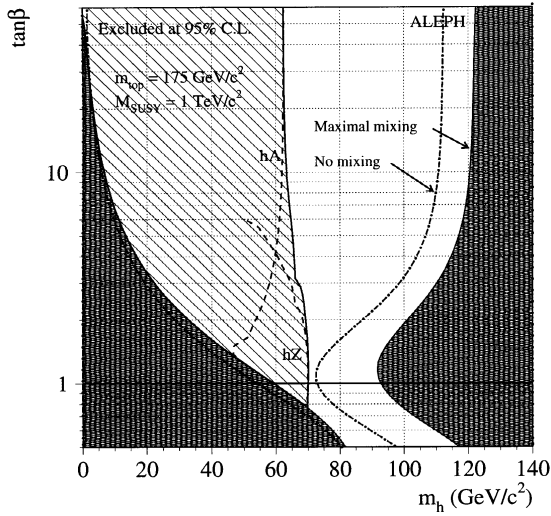}

\caption{The $(m_h, \tan\beta)$ plane in the maximal stop mixing configuration. The dark areas are theoretically disallowed. The hatched area is excluded at 95\% confidence level by the combined search for $e^+e^-\to hZ$ and $e^+e^-\to hA$. The dot-dashed lines show the change in the theoretically excluded region in the no mixing configuration. From Ref.~\protect\cite{aleph}}
\end{figure}

\section{$\tan\beta$ from top Yukawa fixed point}

In a unified MSSM a large top quark Yukawa coupling $(\lambda_t)$ at the GUT scale is needed for radiative electroweak symmetry breaking and for $b$-$\tau$ Yukawa unification\cite{bbo-93,bbop,bagger}. Then $\lambda_t$ approaches an infrared point, giving a relation between $m_t$ and $\tan\beta$ that is independent of the $\lambda_t$ value at $M_{\rm GUT}$. Figure~8 shows the $\lambda_t$ fixed point regions in the $\tan\beta$ vs.\ $m_t^{\rm pole}$ plane. For $m_t = 175$~GeV there are two solutions
\begin{eqnarray}
&&\tan\beta \approx \phantom01.8 \quad\, (\lambda_b^{\rm GUT} \simeq \lambda_\tau^{\rm GUT}) \,,\\
&&\tan\beta \approx 56\qquad (\lambda_t^{\rm GUT} \simeq \lambda_b^{\rm GUT} \simeq \lambda_\tau^{\rm GUT}) \,.
\end{eqnarray}
These fixed point solutions are very attractive theoretically, but it is premature to rule out $\tan\beta$ values between the fixed-point values. The Higgs couplings to fermions and to weak bosons are strongly dependent on $\tan\beta$, as illustrated in Fig.~9. Search strategies must allow for large variations in the possible coupling strengths.

\begin{figure}[h]
\centering
\hspace{0in}\epsfxsize=7.5cm\epsffile{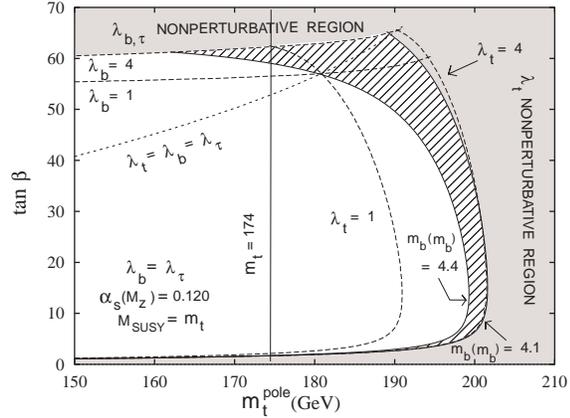}

\caption{Contours of constant $m_b(m_b)$ in the $m_t(m_t), \tan\beta$ plane with contours of constant GUT scale Yukawa couplings.  From Ref.~\protect\cite{bbo-93}.}
\end{figure}

\begin{figure}[t]
\centering
\hspace{0in}\epsfxsize=7cm\epsffile{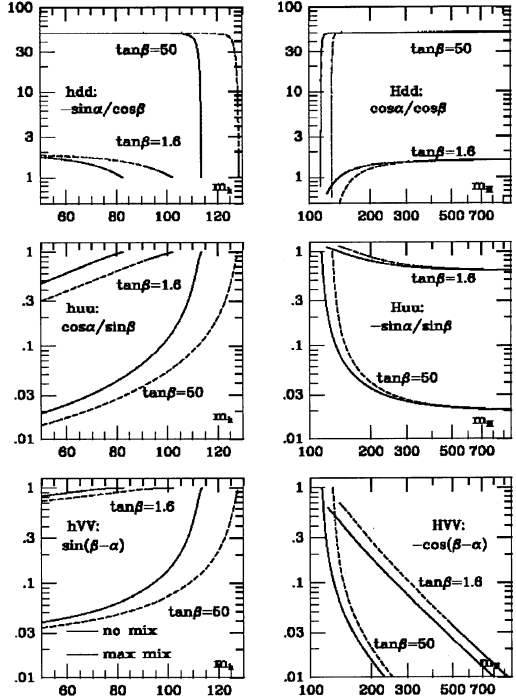}

\caption{MSSM couplings normalized to the SM couplings. From  Ref.~\protect\cite{higgs-lep2}.}
\end{figure}

\section{mSUGRA model}

The existence of a low-energy supersymmetry may be indicated by the successful unification of the gauge couplings. In minimal supergravity (mSUGRA) models, the Higgs boson mass spectrum depends mainly on the universal scalar ($m_0$) and gaugino ($m_{1/2}$) masses at the GUT scale and on $\tan\beta$\cite{ellis}. Figure~10 illustrates representative Higgs masses for the low and high $\tan\beta$ fixed point solutions\cite{dm}. For $\tan\beta\simeq1.8$ the $A^0$, $H^0$, and $H^\pm$ are considerably more massive than $h^0$, whereas for $\tan\beta\simeq50$, the $A^0$, $H^0$, and $H^\pm$ can be comparable in mass to $h^0$. In the low $\tan\beta$ fixed point scenario the $h^0$ has a mass range that is accessible to \mbox{LEP-2}.

There are two interesting $h^0$ coupling regimes, according to whether $A$ is much larger or comparable in mass to the $Z$\cite{habernir}; see Table~3.

\begin{table}
\caption{Two $h^0$ coupling regimes.}
\medskip
\centering
\leavevmode
\epsfxsize=3in\epsffile{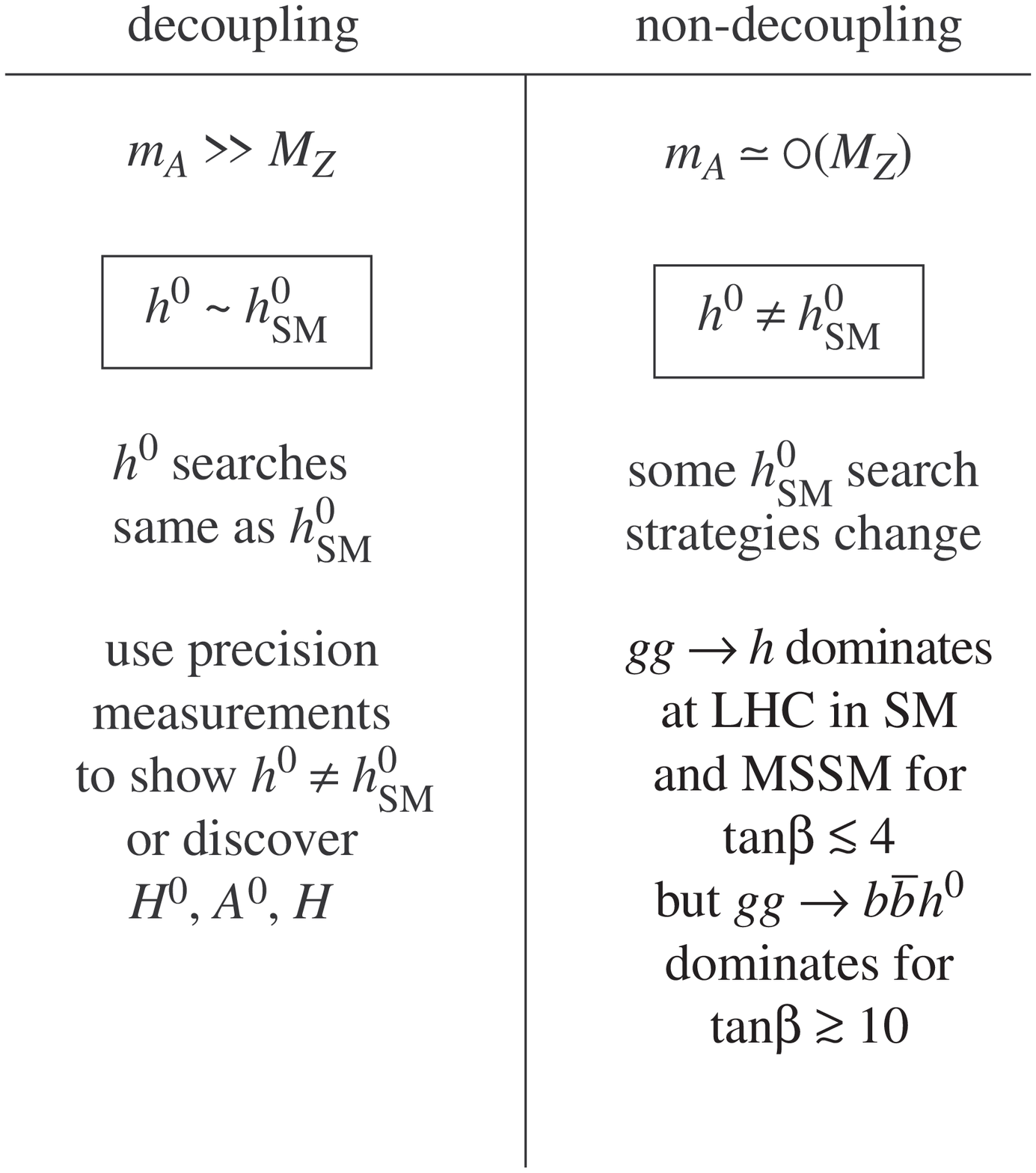}
\end{table}

At the LHC at least one of the MSSM Higgs bosons should be discovered. The coverage of the $( m_A,\tan\beta)$ plane from various channels is shown in Fig.~11, assuming $M_S =1 $~TeV and no squark mixing. However, for some regions of the parameter space only $h$ could be found.

The discovery potential for MSSM Higgs boson at various colliders is summarized in Table~4. For complete coverage of the Higgs states, high energy $e^+e^-$ and $\mu^+\mu^-$ colliders may be necessary if the masses of the neavier states are ${\cal O}$(TeV).

\begin{figure*}[t]
\centering
\hspace{0in}\epsfxsize=10cm\epsffile{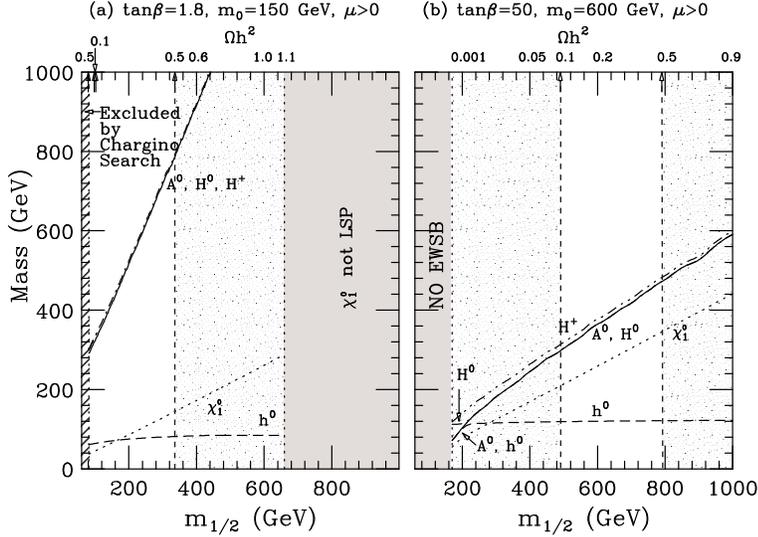}

\caption[]{ The neutralino relic density 
and SUSY Higgs mass spectrum versus $m_{1/2}$ for $\mu > 0$ 
with (a)~$\tan\beta = 1.8$, $m_0 = 150$ GeV and  
(b)~$\tan\beta = 50$, $m_0 = 600$ GeV.
$\tilde{\nu}$ is the lightest scalar neutrino.
The shaded regions denote the parts of the parameter space 
(i)~producing $\Omega_{\chi^0_1} h^2 < 0.1$ or $\Omega_{\chi^0_1} h^2 > 0.5$, 
(ii)~excluded by theoretical requirements, 
or (iii)~excluded by the chargino search at \mbox{LEP-2}. From Ref.~\cite{dm}.
}
\end{figure*}

\begin{figure}[t]
\centering
\hspace{0in}\epsfxsize=7.5cm\epsffile{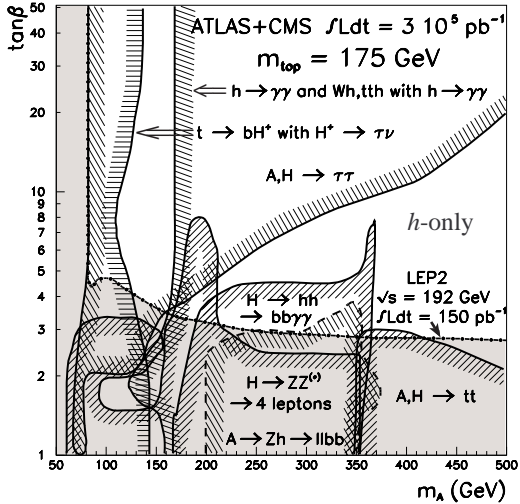}
\caption[]{MSSM Higgs discovery regions $(5\sigma)$ in $(m_A,\tan\beta)$ parameter space for the ATLAS+CMS experiments at the LHC with 300~fb$^{-1}$ per detector. From Ref.~\protect\cite{froid}.}
\end{figure}

\begin{table}
\caption{Discovery potential for MSSM Higgs bosons}
\centering
\hspace{0in}\epsfxsize=7.5cm\epsffile{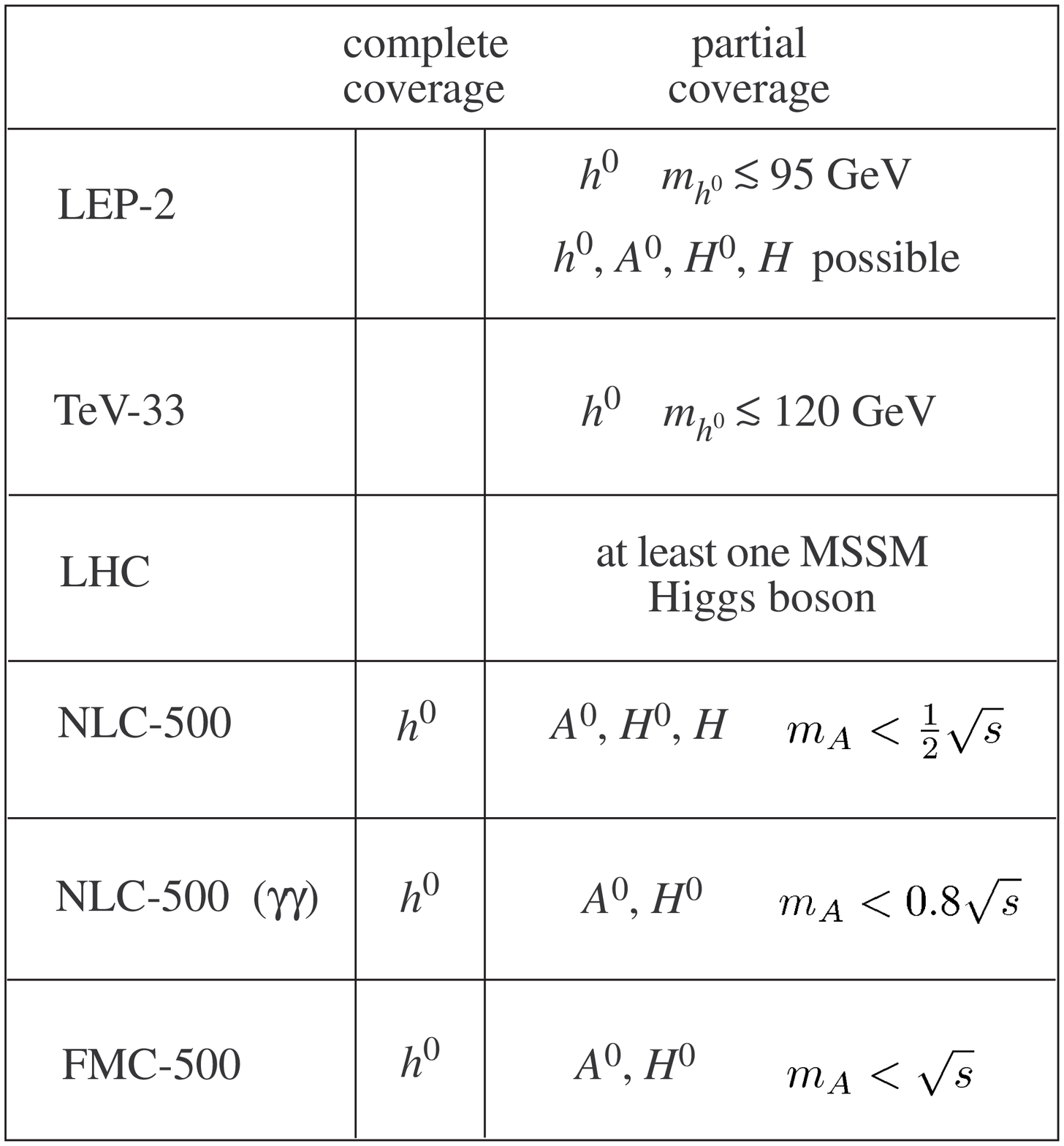}
\end{table}

\section{What we will learn from $h$ searches}

If $m_h$ exceeds 130~GeV, the MSSM is inconsistent. If $m_h$ exceeds 150~GeV, weakly coupled supersymmetry is inconsistent. If $m_h$ is below 130~GeV, the MSSM is viable, but is the Higgs boson the MSSM Higgs? Discovery of only one Higgs boson is insufficient to establish the MSSM. High energy lepton colliders would insure discovery of heavy Higgs bosons. At a 1.5~TeV $e^+e^-$ collider, the $A, H^0, H^\pm$ states could be discovered for $m_A<0.75$~TeV. At a 4~TeV $\mu^+\mu^-$ collider, these Higgs bosons can be found for $m_A<4$~TeV. 

\section{$h^0_{\rm SM}$ or $h^0_{\rm MSSM}$?}

There are two ways to know whether a Higgs boson is a SM or MSSM member. The 
first way is to discover the other MSSM Higgs bosons. The second way is to measure the $h^0$ branching fractions and total width. The enhanced couplings to $b\bar b$ and $\mu^+\mu^-$ of the $h^0_{\rm MSSM}$ compared to $h^0_{\rm SM}$ give MSSM/SM coupling ratios exceeding 1 by ${\sim}10\%$ or more if $m_A \lsim 400$~GeV. The $h^0$ branching fractions for the decays $h^0\to b\bar b, c\bar c, ZZ^*, WW^*, \gamma\gamma$ can be measured to ${\sim}10\%$ accuracy at $e^+e^-$, $\gamma\gamma$ and $\mu^+\mu^-$ colliders\cite{haber,gunion}.

The total widths of the Higgs boson can be another sensitive gauge of the Higgs nature. For example, if $m_A\lsim 150$~GeV, then $\Gamma_{\rm total}(h^0_{\rm MSSM})$ grows proportional to $\tan^2\beta$; for $\tan\beta\sim1.8$, $\Gamma_{\rm total}\sim5$~MeV but for $\tan\beta\sim50$, $\Gamma_{\rm total} \sim 0.5$~GeV. The $s$-channel production of Higgs bosons at a muon collider can be used to scan over the Breit-Wigner resonance profile and directly measure $\Gamma_h$\cite{physrep}.

\section{Post-discovery: establish pedigree of Higgs boson}

Once a Higgs boson is discovered the next step is to measure its properties. Is it a spin zero particle? If $h\to\gamma\gamma$ decay is observed then the spin${}\neq1$ by Yang's theorem. Its spin can be established by measuring angular distributions in $h\to b\bar b$, $h\to e^+e^- e^+e^-$, etc. Is the Higgs boson CP even, odd, or an admixture? The general forms of the couplings to fermions $\bar f(a + \kappa \gamma_5)f$ and $Z$ bosons $g_{\mu\nu} + i\kappa\epsilon_{\mu\nu\alpha\beta} P_1^\alpha P_2^\beta / \Lambda^2$ can be measured from angular correlations\cite{haber}. Are the tree level coupling strengths proportional to $m_f$ and $M_Z^2$ as predicted? Are the loop couplings $gg\to h$ and $h\to \gamma\gamma$ of the strengths predicted from known particles in the loops? The detailed studies required to establish the Higgs boson properties should be easier at $e^+e^-$ and $\mu^+\mu^-$ colliders than at hadron colliders. 

\section{The dark matter card}

With $R$-parity conservation, the lightest supersymmetric particle (LSP) is a natural candidate for the dark matter in the universe. In mSUGRA models the lightest neutralino $\chi_1^0$ is the preferred LSP\cite{ellis}. The LSP relic density $\Omega = \rho/\rho_{\rm critical}$ is governed by the annihilation cross section of LSPs in the early universe\cite{jungman} and the $\chi_1^0\chi_1^0\to\rm Higgs$ and $Z$ resonance poles are very influential\cite{dm,griest}. Hence the relic density depends on the masses and widths of the neutral Higgs bosons. A cosmologically interesting  relic density
\begin{equation}
0.1 \lsim \Omega h^2 \lsim 0.5
\end{equation}
implies significant constraints on the mSUGRA parameters and the Higgs mass spectrum, as shown in Fig.~10. As the cold dark matter component of $\Omega h^2$ becomes more  precisely known over the next few years, the corresponding restrictions on the Higgs mass spectrum will be increasingly interesting.

\section{Conclusions}

Electroweak symmetry breaking and mass generation are mysteries that will be solved by forthcoming collider experiments. Precision electroweak tests already point the smoking gun towards Higgs bosons. It is still a horse race to the discovery line for $h^0$. At LEP-2 the energy is critical. At the Tevatron the luminosity is critical. At the LHC at least one Higgs boson should be found. The genetic traits (couplings, branching fractions, total widths) of the lightest Higgs are critical clues to decide if it is the SM, MSSM, a generic 2-doublet member, or otherwise. The top Yukawa infrared fixed points at $\tan\beta\approx1.8$ and $\tan\beta\approx56$ in the unified MSSM are particularly attractive scenarios; the Higgs boson couplings are very different at low and high $\tan\beta$. Decoupling of the light Higgs boson ($h^0\approx h^0_{\rm SM}$) occurs at large $m_A$; for non-decoupling all MSSM Higgs boson are relatively light. The Next Linear Collider ($e^+e^-$) and First Muon Collider ($\mu^+\mu^-$) are essential for definitive studies of Higgs boson properties (spin, CP, decay widths). The high energy lepton colliders NLC($e^+e^-$)-1.5~TeV and NMC($\mu^+\mu^-$)-4~TeV would provide the energy reach for a high mass Higgs spectrum. Finally, the existence of an interesting cosmological relic density imposes strong constraints on the mSUGRA parameter space and the Higgs boson masses.

\bigskip\noindent{\bf Acknowledgments}:
I thank Chung Kao and Tao Han for valuable advice in the preparation of this talk. This work was supported in part by the U.S.~Department of Energy under Grant No.~DE-FG02-95ER40896 and in part by the University of Wisconsin Research Committee with funds granted by the Wisconsin Alumni Research Foundation.

\end{document}